\documentclass[10pt]{article}

\usepackage{amsmath}

\usepackage{times}

\usepackage{graphicx,color,subfigure}

\title{Tracking hidden objects with a single-photon camera}

\author
{Genevieve Gariepy,$^{1\ast}$ Francesco Tonolini,$^{1}$ Robert Henderson,$^{2}$ \\Jonathan Leach$^{1}$ and Daniele Faccio$^{1\ast}$\\
\\
\normalsize{$^{1}$Institute of Photonics and Quantum Sciences, Heriot-Watt University,}\\
\normalsize{David Brewster Building, Edinburgh, EH14 4AS, UK}\\
\normalsize{$^{2}$Institute for Micro and Nano Systems, University of Edinburgh}\\
\normalsize{Alexander Crum Brown Road, Edinburgh EH9 3FF, UK}\\
\\
\normalsize{$^\ast$To whom correspondence should be addressed; }\\
\normalsize{E-mail: genevieve2.gariepy@gmail.com;  d.faccio@hw.ac.uk.}
}

\date{}

\begin{document}

\baselineskip18pt

\maketitle

\begin{abstract}
The ability to know what is hidden around a corner or behind a wall provides a crucial advantage when physically going around the obstacle is impossible or dangerous. Previous solutions to this challenge were constrained e.g. by their physical size, the requirement of reflective surfaces or long data acquisition times. These impede both the deployment of the technology outside the laboratory and the development of significant advances, such as tracking the movement of large-scale hidden objects. We demonstrate a non-line-of-sight laser ranging technology that relies upon the ability, using only a floor surface, to send and detect light that is scattered around an obstacle. A single-photon avalanche diode (SPAD) camera detects light back-scattered  from a hidden object that can then be located with centimetre precision, simultaneously tracking its movement. This non-line-of-sight laser ranging system is also shown to work at human length scales, paving the way for a variety of real-life situations.  
\end{abstract}

Laser illuminated detection and ranging (LIDAR) is a powerful tool that allows to map the position of objects over large distances by sending a laser signal and recording the time it takes to come back \cite{lidar}. However, the technique is limited to objects that are in the direct line-of-sight. In the past years, technologies have emerged that may overcome this limitation and observe objects that are hidden from view by an obstacle such as a wall, e.g. using radar technology \cite{radar1,radar2} or mirror reflections \cite{mirror}. One promising method for looking around obstacles is actually based on the LIDAR concept. Velten~\textit{et~al.} showed that it is possible to scatter light around an obstacle from a vertically placed wall or door  and then detect the light that is backscattered from behind the obstacle  \cite{Velten2012,Gupta2012}. Using a streak camera with picosecond temporal resolution, they could reconstruct full three-dimensional images of an object hidden from the direct line-of-sight. This method leads to accurate object reconstruction but requires long acquisition times, raster-scanning of the laser beam and the presence of a wall or door to act as a scattering surface. A different approach based on a technique for imaging through opaque barriers \cite{Mosk1, Mosk2, Mosk3}, was proposed by Katz \textit{et al.}, where light originating directly from the region outside the line-of-sight (behind the obstacle)  produces a speckle pattern in reflection from a diffusing wall that does lie within the line-of-sight of the camera and can then be post-processed to retrieve the shape of the illuminated object \cite{Ori2012,Ori2014}. This method is very fast as it requires only one image, but requires a light source that is independently placed behind the obstacle so as to directly illuminate the object.  Notwithstanding these ingenious solutions, locating the position of a hidden object and monitoring its movement in real time with a technology that is easily transposable to real-life applications and to the human scale, remains to date a major challenge.

\paragraph{} Our solution to this challenge is based on a single-photon avalanche diode (SPAD) camera in a LIDAR-like system. This technology offers many advantages: it has extremely high sensitivity and can detect single photons and has precise timing resolution ($\sim50$ ps) \cite{Gariepy2015,Richardson1,Richardson2,Charbon,TCSPC,TCSPC2} (see Supplementary Information). The high sensitivity of the camera is crucial as it allows extremely short acquisition times, which in turn allows one to locate hidden objects in real-time, i.e. on time scales sufficiently short to be able to track their movement. The high temporal resolution of the camera relies on the fact that each individual pixel is operated in ``time-correlated-single-photon-counting" (TCPSC) mode \cite{TCSPC,TCSPC2}, i.e. acquisition is performed by measuring the arrival times of single photons and accumulating the measurement over many separate individual laser pulses that illuminate the scene. Finally, we introduce an additional solution with respect to previous work, which although very simple, we believe marks a significant step forward in removing any physical constraints on the surrounding environment for the deployment of the technology in real-world situations: we use the floor as a scattering and imaging surface, thus eliminating the need for a close-by wall, door or highly reflective surface. 

\paragraph{}In the following, we first show that we can locate the position of an object hidden behind a wall with centimeter precision, without the need for pre-acquiring a background  in the absence of the object. Due to the sensitivity of the camera and the fact that no scanning of the laser is needed, the data acquisition for a fixed position of the object takes only three seconds. We then show that real-time acquisition is possible for an object moving at a few centimeters per second. Finally, we discuss the potential of this technology for real-world applications. 

\begin{figure} [t]
\centering
\includegraphics[width=0.75\textwidth]{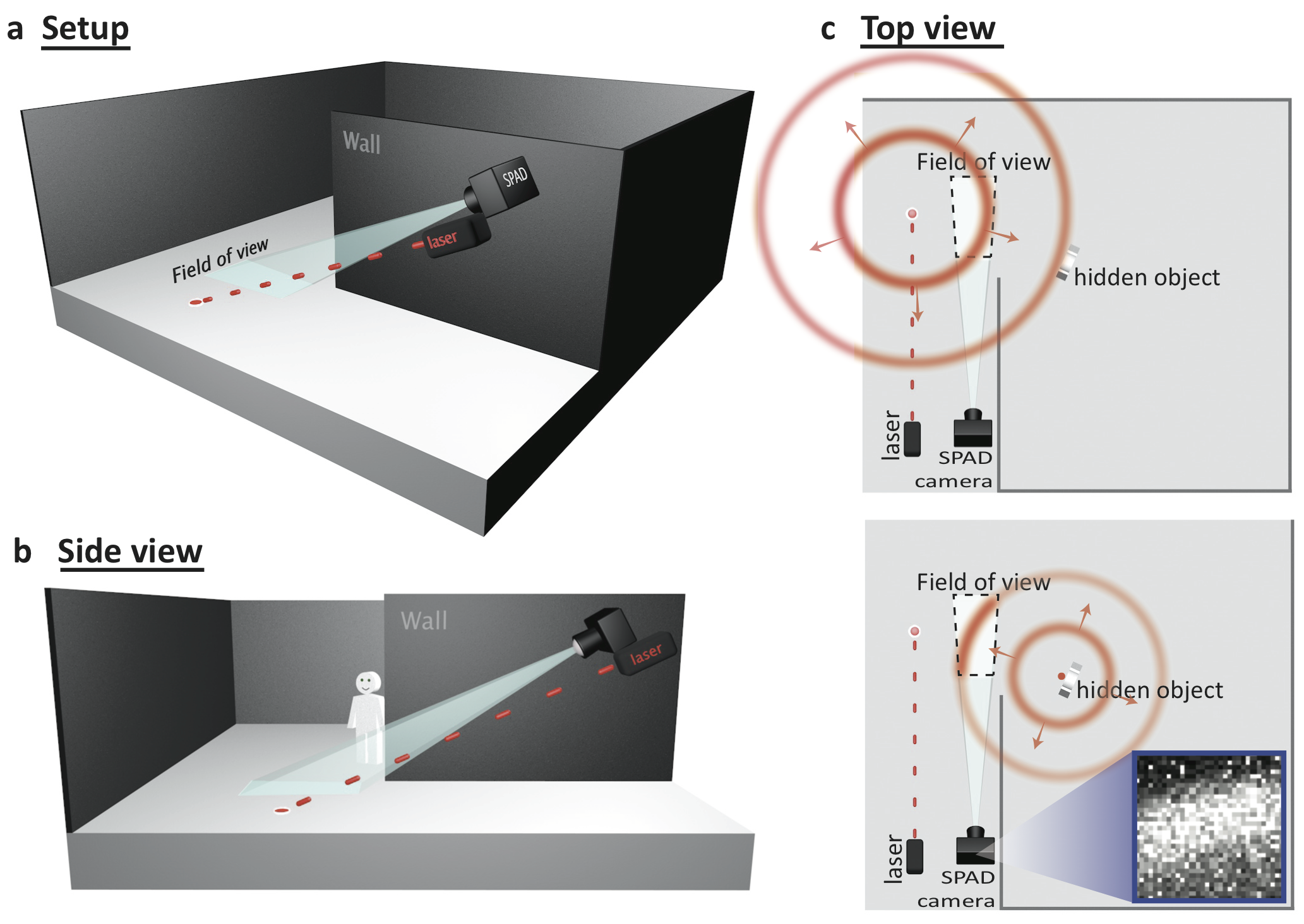}
\caption{\label{figure1}\textbf{Looking around a corner.} Our setup re-creates, at a $\sim5$x reduced scale, a situation where a person is hidden from view by a wall or an obstacle. a) The camera is positioned on the side of the wall and is looking down at the floor: it cannot see what is behind the wall but its field of view is placed beyond the end of the obstacle. b) A side view shows that the target is hidden behind the wall. In order to see the hidden target around the corner, laser pulses are sent to the floor. c) The light then scatters off the floor and propagates as a spherical wave behind the obstacle, reaching the hidden object. This light is then in turn scattered back into the field of view of the camera. The SPAD camera records both spatial and temporal information on the propagating spherical light wave as it passes through the field of view, creating an elliptical pattern where it intercepts the floor. An example of the spatially resolved raw data, as recorded by the camera for a fixed time frame as the ellipse passes in the field of view is shown in the inset. }
\end{figure}

\paragraph{} The experiments were performed in the laboratory where we re-create conditions of a person moving behind a wall, as shown in figure \ref{figure1}, albeit scaled down from real-size by factor $\sim5\times$. The ``target" we wish to track is therefore a human form cut in a piece of foam,  that is 30 cm high, 10 cm wide and 4 cm thick. The SPAD camera is placed beside a wall that hides the target from its view. The target is positioned roughly one meter away from the camera (Fig.~\ref{figure1}a,b). As in many real-life situations, there is not always a conveniently placed wall, door or window that can be used as a reflective surface to send and collect light: we rely only on the presence of the floor, in this case, a piece of white cardboard. Other floor surfaces were tested e.g. black cardboard leading to very similar results albeit requiring $\sim$5$\times$ longer acquisition times. The camera is therefore imaging a patch of the floor that is just beyond the edge of the obscuring wall. We then send a train of femtosecond laser pulses (800 nm wavelength, 67 MHz repetition rate, 0.67 W average power) on the floor, 15~cm to the left of the field of view of the camera. Light scatters from this point into a spherical wave and propagates behind the wall (see Fig.~\ref{figure1}c). Part of the scattered light reaches the hidden target, which then scatters light back into the field of view. This light is detected by the camera and is used to reconstruct the position of the target. We underline here that the camera has an impulse response function of $\sim110$ ps, corresponding to a spatial (depth) resolution of a few centimeters, i.e. of the same order of magnitude of our target. On the one hand this implies that we are not able to directly reconstruct a 3D image of the target as performed in Ref.~\cite{Velten2012}, but on the other hand it allows us to approximate the  back scattering as a single spherical wave originating from the target. 

\paragraph{} The SPAD camera is sensitive enough to detect light after multiple scattering events in a few seconds of acquisition time. It also yields temporal information, i.e. it can record the temporal evolution and details of the backscattered light with $\sim110$ ps temporal and 32x32 pixel spatial resolution (see Supplementary Information). In Fig.~\ref{figure1}c, we show an example of a single time frame of such a recording (raw acquired data) where the back reflected spherical wave is clearly visible, frozen in time by the camera as it sweeps across the field of view.  The target location is then retrieved by utilising the fact that: (i) the time it takes for the light to propagate from the laser to the object and back, similarly to a LIDAR system, gives information about the object's distance and (ii) the curvature and direction with which the spherical wavefront propagates across the camera field-of-view provides information on the target position.

\begin{figure}[h]
\centering
\includegraphics[width=0.85\textwidth]{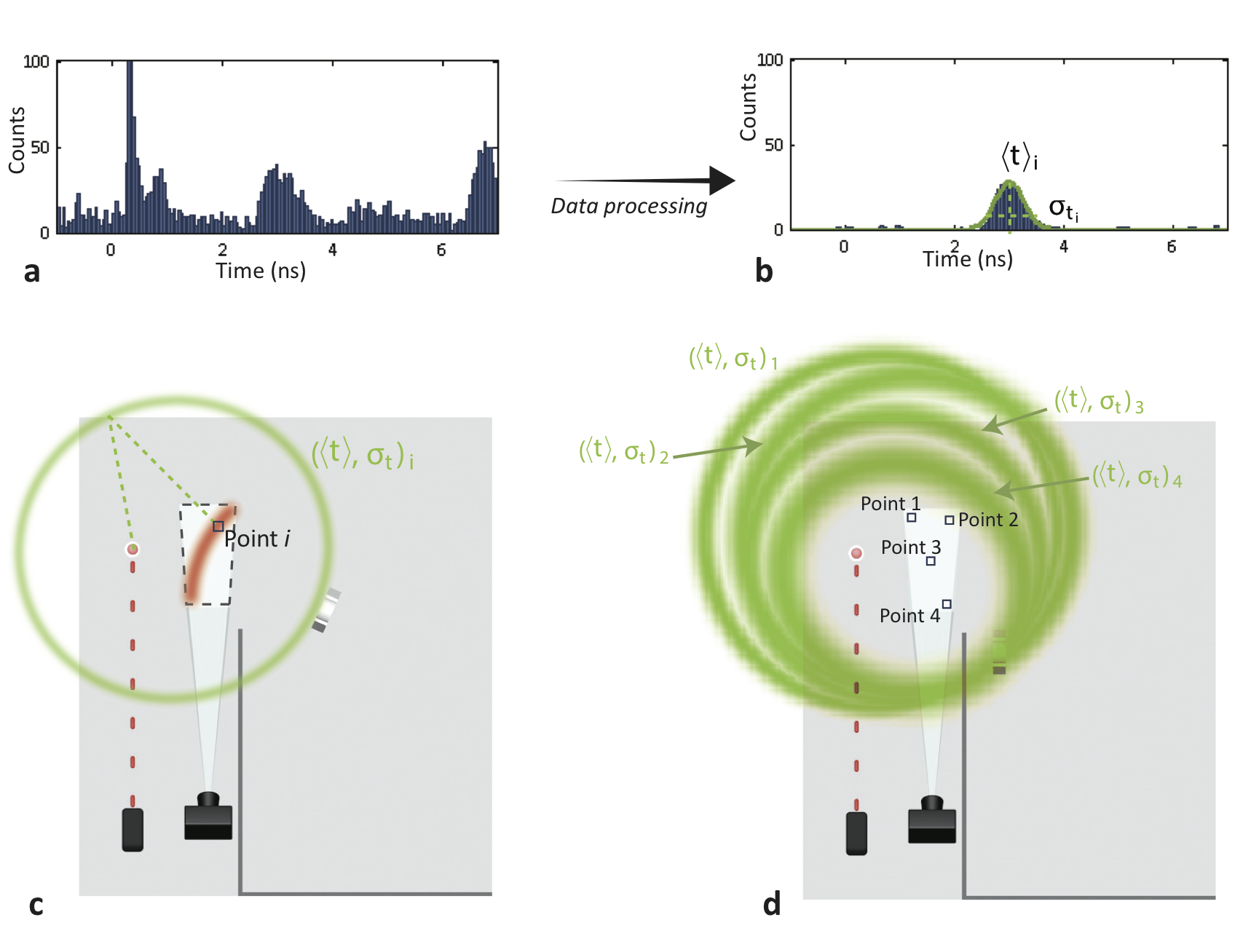}
\caption{\label{figure2} \textbf{Retrieving a hidden object's position.} a) A histogram of photons arrival times is recorded for every pixel (here for pixel $i$ as indicated in \textit{c}). This experimental histogram contains signals both from the target and unwanted sources. b) Background subtraction and data processing allows to isolate the signal from the target and fit a Gaussian to its peak, centered at $\langle t\rangle_i$ with a standard deviation of $\sigma_{t_i}$.  c) The time of arrival $\langle t\rangle_i$ is used to trace an ellipse of possible positions of the target which would lead to a signal at this time.  d) Ellipses calculated from different pixel (experimental data) give slightly displaced probability distribution that intercepts at a given point. Multiplying these probability distributions (also with all other similar distributions from all 1024 pixels of the camera) provides an estimate of the target location.}
\end{figure}

\paragraph{} \textit{Target position retrieval.} The target-position retrieval algorithm therefore relies on both the temporal and spatial information recorded by the SPAD camera. Every pixel $i$ of the 32x32-pixel camera, corresponding to a position $\vec{r}_i = (x_i, y_i)$ in the field of view, records a histogram of photon arrival times (see Fig.~\ref{figure2}a). The temporal histogram of a given pixel will show the temporal evolution signal corresponding to light scattered back from the target, but also from unwanted sources from the environment such as the walls and the ceiling. The first step is to isolate the signal of interest coming from the target alone. This can be achieved by simply acquiring a background signal in the absence of the target. Although this proves to be the most effective method, if we are interested in tracking non-cooperative moving targets, we may also use a self-referenced background subtraction scheme. Indeed, by acquiring data with the target at different positions, we can distinguish the signal that is not changing at each acquisition (i.e. therefore generated by the static sources) and the signal that is changing (i.e. therefore generated by the target). A median of the temporal histograms for each pixel proves to be a very a good approximation of the background signal \cite{book1, back1,back2} and allows to effectively isolate the signal generated in reflection from the target alone (see Supplementary Information). 

\paragraph{}Once the target signal is isolated, we proceed similarly to standard time-of-flight measurements and fit a Gaussian function to the temporal histogram recorded at each pixel, as shown in Fig.~\ref{figure2}b \cite{Gerald, ranging1, ranging2, ranging3}. For a pixel $i$, the peak position of the Gaussian fit $\langle t\rangle_i$ is used to determine the total photon flight time, with an uncertainty that is taken to be the Gaussian standard deviation $\sigma_{t_i}$. The time $\langle t\rangle_i$  measures the light travel-time from the moment the laser hits the ground, scatters to an object at a point $\vec{r}_o = (x_o,y_o)$ and scatters back to the specific point $\vec{r}_i$ in the field of view of the camera. However, there is a whole ensemble of points or target locations $\vec{r}_o$ that satisfy this condition. More precisely these points form  an ellipse defined by $|\vec{r}_o-\vec{r}_l|+|\vec{r}_o-\vec{r}_i| = \langle t\rangle_i\times c$, where $|\vec{r}_o-\vec{r}_l|$ and $|\vec{r}_o-\vec{r}_i|$ are the distances from the laser point $\vec{r}_l$ on the floor to the target and from the target to the point $\vec{r}_i$, respectively, as illustrated in Fig.~\ref{figure2}c.  This ellipse represents a probability distribution for the position of the hidden object with uncertainty $\sigma_{t_i}$:

\begin{equation} P_i^{\text{ellipse}}(\vec{r}_o) \propto \exp{-\frac{(\varepsilon/c-\langle t\rangle_i)^2}{2\sigma_{t_i}^2}} \end{equation}

 \noindent \paragraph{} where $\varepsilon$ is the ellipsoidal coordinate $\varepsilon = |\vec{r}_o-\vec{r}_l|+|\vec{r}_o-\vec{r}_i|$.  We therefore calculate the probability distributions $P_i^{\text{ellipse}}(\vec{r}_o)$ for every pixel $i$ of the field of view. Figure~\ref{figure2}d shows as an example, four of the $P_i^{\text{ellipse}}(\vec{r}_o)$ probabilities calculated from experimental data, correponding to four different pixels indicated in the figure. The area where the ellipses overlap indicates the region of highest probability for the target location. In order to retrieve the target's position, we calculate the joint probability density by multiplying the probability densities from all 1024 camera pixels: 

\begin{equation} P(\vec{r}_o) =N \prod_{i=1}^{1024}P_i(\vec{r}_o). \end{equation}

\paragraph{} $P(\vec{r}_o)$ determines the overall probability distribution of the location of the target, and $N$ is a normalisation constant. A complete mathematical development and details about the form of $P_i(\vec{r}_o)$ are given in Supplementary Information. 

\begin{figure} [hb!]
\centering
\includegraphics[width=0.8\textwidth]{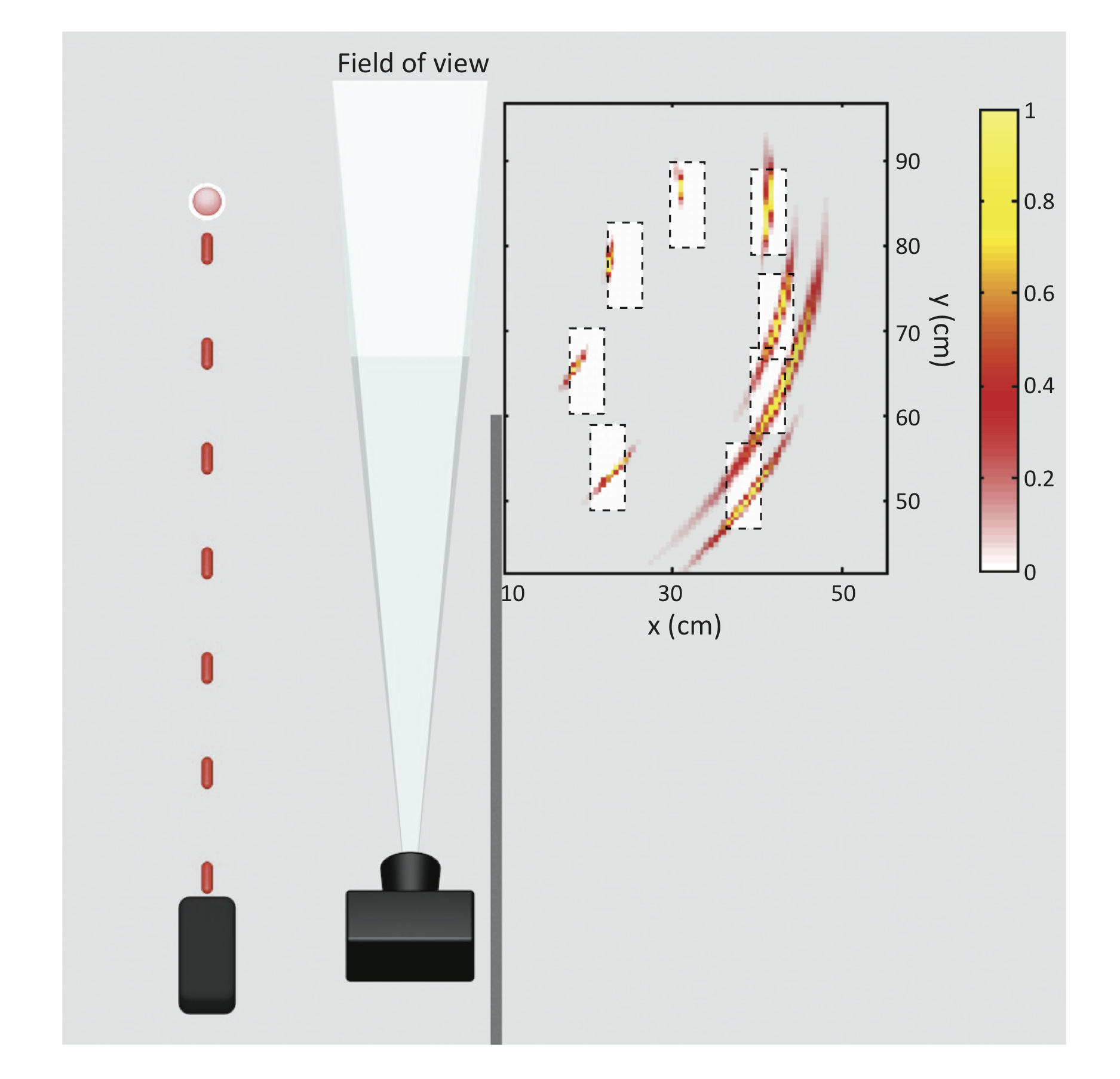}
\caption{\label{figure3}\textbf{Experimental results of hidden object's position retrieval.} Experimental layout and results showing the retrieved locations for eight distinct positions of the target, approximately one meter away from the camera (distances indicated in the figure are measured from the camera). The coloured areas in the graph indicate the joint propability distribution for the target location whose actual positions are shown as white rectangles.}
\end{figure}

\paragraph{} \textit{Results.} In a first experiment, we place the target at eight distinct positions and acquire data for three seconds at each position. We then use the  algorithm detailed above to retrieve the position of our target. The results are shown in Fig.~\ref{figure3}, where the probability density $P(\vec{r}_o)$ is shown for the eight positions of the target. The figure shows, to scale, the relative positons of the laser illumination spot on the floor, the camera and its field of view, together with the actual positions of the target superimposed on the joint probability distributions in color scale (each $P(\vec{r}_o)$ peak value is individually normalised to one). The method provides an accurate retrieval of the target's position with a precision of approximately $\pm5$~mm in $x$ and $\pm1.5$~cm in $y$, corresponding to $\sim10\%$ uncertainty with respect to the target's size in both directions. We note that the closer positions unavoidably have a lower uncertainty that is related to the higher overall light signal.  The results also show that we are able to retrieve target positions when it is not only hidden from view but is actually physically receded behind the end of the wall.

\begin{figure} [h!]
\centering
\includegraphics[width=0.6\textwidth]{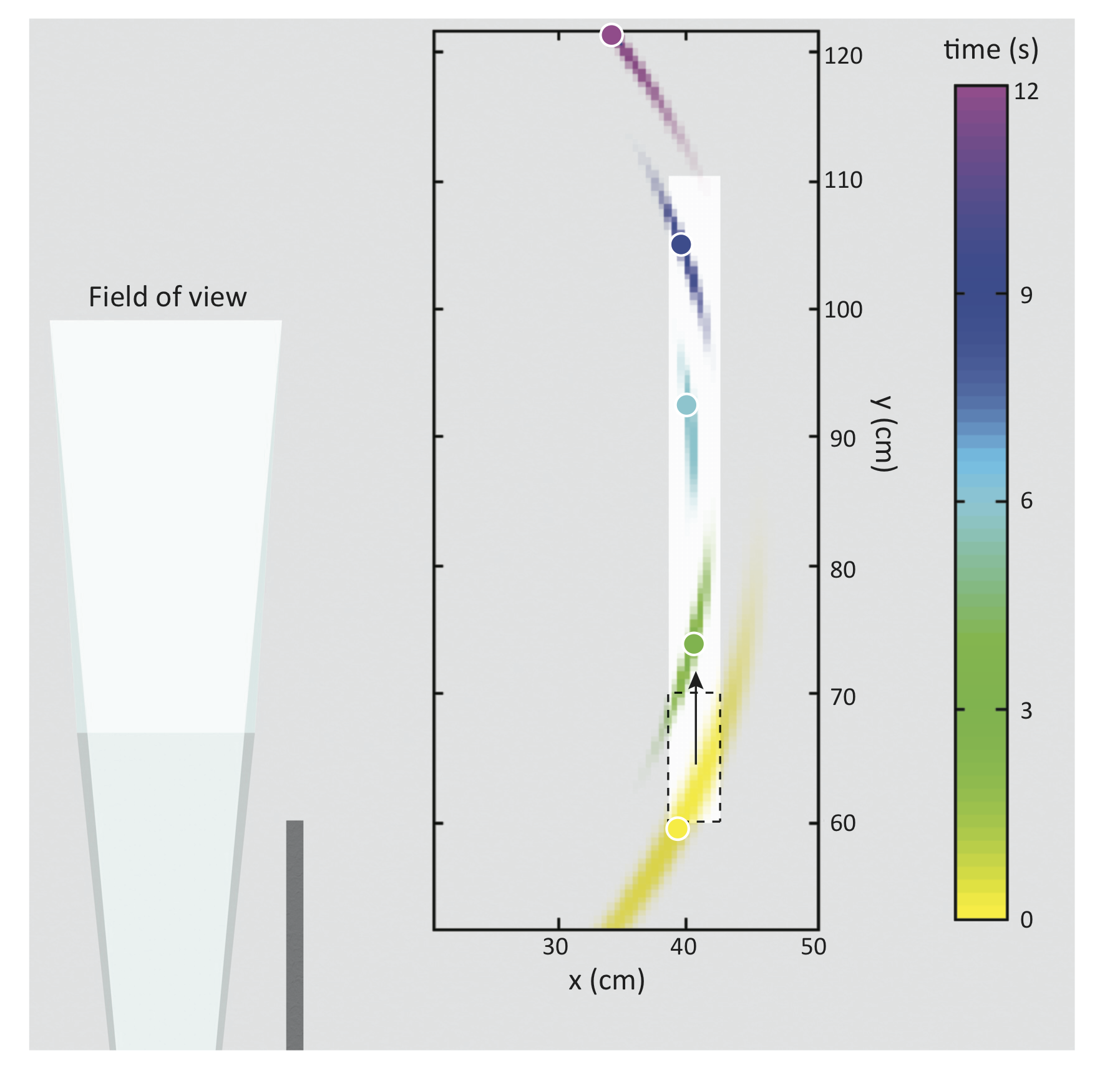} 
\caption{\label{figure4} \textbf{Non-line-of-sight tracking of a moving target.} Distances in the graph are measured from the camera position. The object is moving in a straight line along the $y$-direction, from bottom to top (as represented by the dashed rectangle and the arrow), at a speed of 2.8 cm/s. The coloured areas represent the retrieved joint probability distribtions: the point of highest probability, indicating the estimated target location, is highlighted  with a filled circle.  The colours correspond to different times, as indicated in the colorbar: successive measurements are each separated by  3 second intervals, i.e. the data acquisition time as explained in the text.  }
\end{figure}

\paragraph{} The data for each target position is acquired in three seconds and this time is currently only limited by the USB2.0 data download rate from the camera to the computer. Nevertheless, this is still sufficiently fast to enable data acquisition in real-time, i.e. on a time scale short enough to track a hidden object that is moving. To show this, we place the target on a track directed along the  $y$-direction. Although the target is moving continuously, we are recording one position  every 3 seconds, and we thus retrieve a discrete set of locations, each of which represents the  average position of the target during the acquisition time. Figure~\ref{figure4} shows an example of the system tracking a moving target where the acquisition ``start'' times (each separated by 3 seconds) is encoded in color. The target in this measurement was moving at a speed of 2.8 cm/s and could be accurately followed by the camera. Data was also recorded at different speeds and for other $x$ positions of the track: at 3.9 cm/s and 5.3 cm/s for tracks positioned at $x=20$~cm and $30$~cm, respectively. The results for these other sets of data are shown in the Supplementary Information.

\paragraph{} SPAD detectors originally developed as single pixel elements, are gradually becoming widely available as focal plane arrays. The single photon sensitivity and picosecond temporal resolution provide a unique tool for fundamental studies \cite{Gariepy2015} and our results show that they can also enable real-time non-line-of-sight ranging of a moving target. The time for the actual data acquisition and processing for each position is currently around 3 seconds but we estimate that this can be reduced by an order of magnitude by improving the data download rate and/or directly processing the data on chip.  The measurements also demonstrate a series of solutions that will allow deployment in real-life situations, e.g. self-referenced background subtraction in the presence of moving targets and only the requirement of a floor from where to reflect and then detect the laser pulse signals.  The current limitation in tracking speed is the data acquisition time, i.e. we can reliably locate the position of a target if this moves by less that its own size during the acquisition time. This corresponds to tracking speeds of a few km/h for human-sized objects and, as discussed above, can be improved by a factor 10$\times$.  We also performed preliminary measurements in which we investigated performance at the human-scale, i.e. we tracked in real time the movement of a researcher as she moved in the room, hidden from the direct line-of-sight of the camera. The system continues to provide reliable results, thus paving the way for a number of applications e.g. in surveillance or rescue scenarios.

\clearpage

\renewcommand{\theequation}{S\arabic{equation}}
\renewcommand{\thetable}{S\arabic{table}}
\renewcommand{\thefigure}{S\arabic{figure}}
\renewcommand{\thesection}{S\arabic{section}}

\topmargin 0.0cm
\oddsidemargin 0.2cm
\textwidth 16cm 
\textheight 21cm
\footskip 1.0cm

\begin{center}\huge{\textbf{Supplementary Material} }\end{center}

\vspace{15pt}

\section{Materials and Methods}
\paragraph{} The laser we use in our non-light-of-sight laser ranging system is a femtosecond oscillator that emits pulses of 10~nJ energy and 10~fs duration at a 67 MHz repetition rate. A small portion of the laser (8$\%$ reflection) is sent to an optical constant fraction discriminator (OCF) that generates a TTL signal then sent to the camera to synchronise the acquisition to the propagation of the laser pulses. We note that the system has been tested with different laser specifications, e.g. light-in-flight with the same SPAD camera was demonstrated using a portable micro-chip laser, with 4 kHz repetition rate (ref. 12 in main text).

\paragraph{} The camera is a 32x32-pixel array of Si CMOS single-photon avalanche diodes (SPAD) that are individually operated in time-correlated single-photon counting (TCSPC) mode: every time a photon is detected by a pixel, the time difference between its arrival and the arrival of the TTL trigger from the OCF is measured and stored in the time histogram. Each histogram has 1024 time pixels with a time-bin of 45.5 ps. The time resolution is limited by the electronic jitter of the system, which is $\sim$110 ps. A standard Nikon-mount lens is attached to the camera (Samyang, 8 mm focal length, F3.5).

\paragraph{} The histograms are recorded over 10,000 laser pulses and the camera is operated at its minimum operating exposure time of 300$\mu$s, so that each acquisition takes 3 seconds. The operating frame rate is only limited by the camera's USB connection to the computer, so that the minimum exposure time is currently 300$\mu$s (frame rate of 3 kHz). The next generation of SPAD camera will be implemented with USB3.0 which will allow to reach higher operating rates, up to the limit of 1 MHz set by the camera's internal functioning. The limit at which data can be acquired will then be set by the amount of signal scattered back to the field of view of the camera, but we expect to be able to record data for one position within less than a second. 

\section{Background subtraction and data processing}

\begin{figure} [h]
\centering
\includegraphics[width=0.7\textwidth]{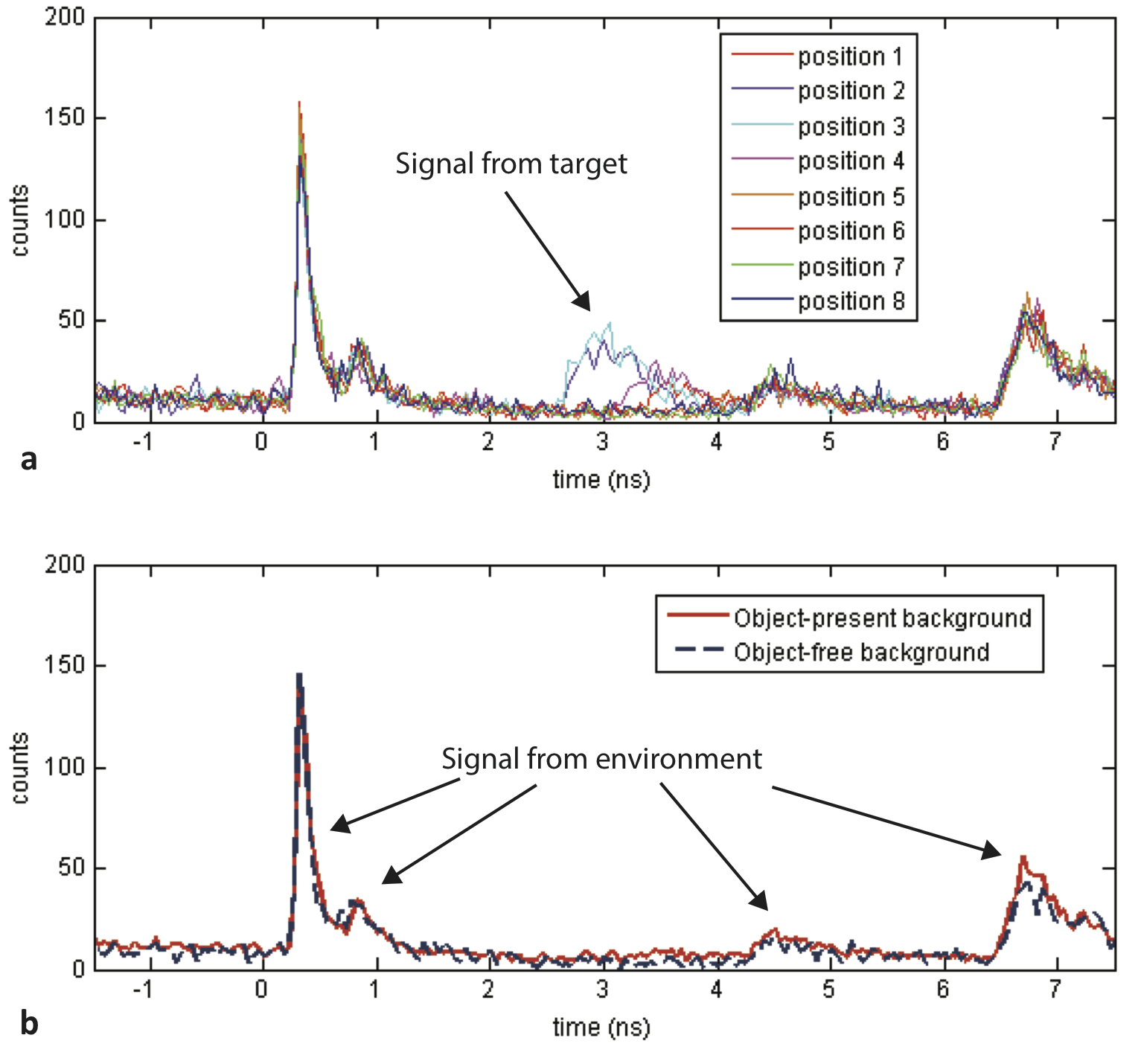}
\caption{\label{background}\textbf{Background subtraction.} To avoid acquiring a background in controlled conditions, where we know the target is not present, we can rely on a few acquisitions where the target is at different positions. a) In the eight histograms shown here, the target was positioned at different places leading to a variation in its detected signal. b) The background calculated from the median of the histograms in \textit{a} (object-present) is very close to the background acquired without the target present (object-free). }
\end{figure}

\paragraph{} As explained in the text, in a case where an object-free background is impractical or impossible to acquire, the median of histograms recorded at different times can be used to estimate the background. We show in figure \ref{background} an example of a background calculated from the median of eight recorded positions. Figure \ref{background}a shows the histograms of a given pixel for the eight positions of the target mentioned in the main text. We can see that the peaks coming from the target are shifting in time when the target changes position, between 2.5~ns and 4.5~ns. The rest of the signal is coming from fixed objects: the walls and the ceiling. 

\paragraph{}At each point in time, we record the median value of the 8 positions. Figure \ref{background}b shows the result of the median-calculated background (object-present background) compared to the object-free background. It gives a fairly good approximation of the background, especially for the peaks coming from fixed object.  

\begin{figure}  [h]
\centering
\includegraphics[width=0.7\textwidth]{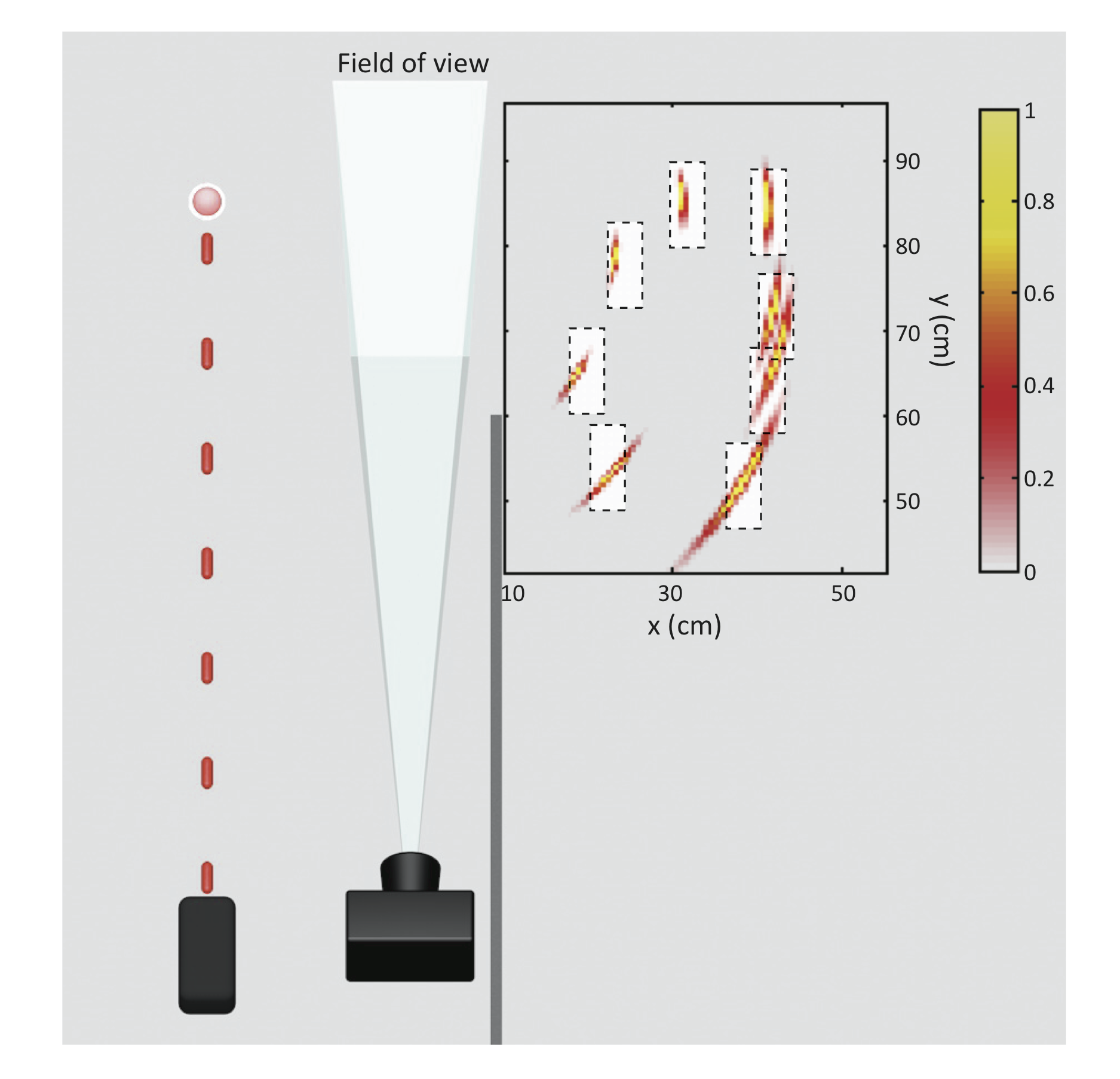}
\caption{\label{locating}\textbf{Locating a target using a pre-acquired background.} We show here the probability densities of the retrieved positions mentioned in figure 3 of the main text. The data was here processed using an object-free background, and it yields to slightly more accurate and precise results than using the object-present background for positions farther away from the camera.}
\end{figure}

\paragraph{}When using this technology in a situation where it is possible to acquire an object-free background, this yields slightly more accurate results. Figure \ref{locating} shows the retrieved position for the experiment mentioned in figure 3 of the main text, but using the object-free background.

\paragraph{} In a real-life implementation, the object-present background can be calculated with the first few acquisitions. If there is no object moving, but a target is present, this algorithm will fail to detect the target; but as soon as the target starts moving, it will take around 15 seconds ($\sim$5 acquisitions) to record a background and begin to accurately locate the target. The best approximation of a background will come from a target that is moving in both the $x$ and $y$ directions. However, we underline that during this initialisation period of 15 seconds, the camera is acquiring data which does provide information regarding the movement of the target: this data is simply less accurate with respect to the data acquired at later times, but nonetheless indicates target movement.

\paragraph{}To retrieve the position of the hidden target, we also need to know the position $\vec{r}_i$ of pixel $i$ on the floor.  The camera is looking down at the floor (it is located at a 46 cm height above the floor and is imaging a region, on the floor, located  67 cm ahead), but still records a 32x32 square image. This image actually corresponds to a field of view that is trapezoidal and stretched both spatially and temporally, with respect to a squared field of view perpendicular to the line of sight of the camera. To correctly retrieve the positions of a hidden target, we need to first determine the actual position ($\vec{r}_i=x_{i},y_{i}$) of each pixel of the camera.  To do so, we measure the dimensions of the field of view and its distance to the camera to reconstruct the actual shape of the field of view on the floor. In a real-life application, knowing the height and angle of the camera with respect to the floor will be sufficient to determine the geometry of its field of view and therefore know the positions ($x_{i},y_{i}$) of each pixel in the field of view.  We also correct for temporal distortion of the recorded data: because the field of view is not perpendicular to the camera's line of sight, photons recorded at the top of the field of view take longer to reach the camera than the ones coming from the bottom. Again, knowing the geometry of our imaging system, we correct the measured $\langle t \rangle_i$ accordingly. We then use the values of $\vec{r}_i$, $\langle t\rangle_i$ and $\sigma_{t_i}$ to retrieve the position of a hidden target, as explained in the next section.

\section{Position retrieval}

\paragraph{} The mathematical model we develop for locating objects around corners relies on the time histograms recorded by each pixel of the SPAD camera. These histograms give a probability distribution of arrival times for photons scattered by a hidden target. This time distribution can be mapped into a probability density in space for the target position. A spatial probability density is calculated for each pixel, and the product of all probabilities constitutes the joint probability density of finding the target at a specific position in space. 

\begin{figure}[h]
\centering
\includegraphics[width=0.5\textwidth]{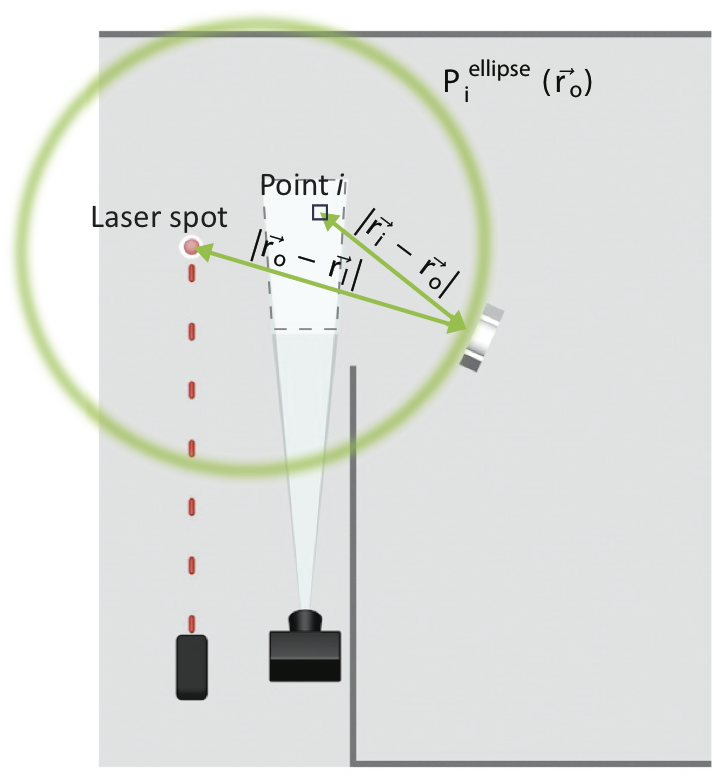}
\caption{\label{variables}\textbf{Retrieving the position of a hidden target} For each pixel $i$, a signal is detected at time $t_i$. $t_i$ is the time it takes for a pulse to propagate from the laser spot to the object ($|\vec{r}_o-\vec{r}_l|$) and from the object to the pixel ($|\vec{r}_{i}-\vec{r}_o|$).  We determine a probability density in space $P_i^{\text{ellipse}}$ of where the object can be located to send a signal at pixel $i$ at time $t_i$.}
\end{figure}

\paragraph{}We first treat each pixel independently. The scattered light from an object at position $\vec{r}_o$ is detected at pixel $i$ (located at position $\vec{r}_{i}$) at  time $t_i$.   The arrival time $t_i$ corresponds to the time the light takes to propagate (at speed $c$) from the laser position on the floor ($\vec{r}_l$) to the object and  then from the object to the pixel:

\begin{equation} c\times t_i = |\vec{r}_o-\vec{r}_l|+|\vec{r}_{i}-\vec{r}_o| \label{stcompress4} \end{equation}
Figure \ref{variables} illustrates the variables and relevant distances used in the calculation. Solving equation \ref{stcompress4} for the object position $\vec{r}_o$ gives infinitely many solutions lying on the surface of an ellipsoid defined by foci $\vec{r}_l$ and $\vec{r}_{i}$:  the possible positions in space that can generate a signal at pixel $i$ at time $t_i$ lie on an ellipsoidal  surface with evenly distributed probability. In the absence of any uncertainties in the measured signals, the resulting probability density of the object's location $P_i^{\text{ellipse}}(\vec{r}_o)$ calculated from the data collected by pixel $i$ is therefore given by
\begin{align}\label{stcompress5}
P_i^{\text{ellipse}}(\vec{r}_o) \propto \left\{
  \begin{array}{lr}
    1 \quad {\rm if} \quad |\vec{r}_o-\vec{r}_l|+|\vec{r}_{i}-\vec{r}_o| = c\times t_i \\
    0 \quad {\rm otherwise}.
  \end{array}
\right.
\end{align}

We write here with a proportional sign and will treat the normalisation issue ($\int_S P_i^{\text{ellipse}}(\vec{r}_o) d \vec{r}_o = 1$) at the end of the development.
The function above can be represented in a  simpler form by using ellipsoidal coordinates with foci $\vec{r}_l$ and $\vec{r}_{i}$
\begin{align}\label{stcompress6}
\begin{array}{lr}
P_i^{\text{ellipse}}(\vec{r}_o) \propto \delta(\varepsilon-ct_i), \quad  {\rm where} \quad \varepsilon = |\vec{r}_o-\vec{r}_l|+|\vec{r}_{i}-\vec{r}_o|.
\end{array}
\end{align}

\paragraph{}In any real implementation, the histogram $h_i(t)$ recorded by any given pixel $i$ will contain an uncertainty on the arrival time of the signal. As a result, the probability density $P_i^{\text{ellipse}}(\vec{r}_o)$ will no longer be a uniform ellipsoid, and the uncertainty in the time histogram can be mapped onto the spatial probability density as:
\begin{align}\label{stcompress8}
P_i^{\text{ellipse}}(\vec{r}_o) \propto \int_{-\infty}^{\infty} \delta(\varepsilon-ct) f_i(t) dt = f_i\left(\frac{\varepsilon}{c}\right).
\end{align}

\paragraph{}In our experiments, the signal recorded in $h_i(t)$ has a Gaussian form with a standard deviation of $\sigma_{t_i}$. The uncertainty is originating from different sources, for example the jitter on the system and the finite size of the target.  As a result of the Gaussian form of the recorded signal,  the general expression of the probability density $P_i^{\text{ellipse}}(\vec{r}_o)$ expressed in \ref{stcompress8} then becomes

\begin{align}\label{stcompress20}
P_i^{\text{ellipse}}(\vec{r}_o) \propto \exp\left[-\frac{( \varepsilon/c-\langle t\rangle_i)^2}{2 \sigma_{t_i}^2}\right].
\end{align}

Here $\langle t\rangle_i$ is the mean arrival time registered by the pixel $i$ and $\sigma_{t_i}$ is its standard deviation. Figure \ref{variables} shows an example of the spatial distribution of $P_i^{\text{ellipse}}(\vec{r}_o)$.

\paragraph{} In our setup we aim to locate the object in $x$ and $y$, on the plane of the floor, making the assumption that such an object is not moving considerably in the vertical direction. We therefore set a two dimensional search space at a given height close to ground level and $\vec{r}_o$ takes the form $\vec{r}_o=(x_o, y_o)$. We take the height of search to be the height of the chest of our foam mannequin $z_o=17$ cm, as this is the region of highest reflectivity. In a real-life implementation, the height can be appropriately estimated based on the type of target we wish to track or locate. An error $\Delta z$ in estimating this height will result in the worst case in an error $\Delta\vec{r}_o$ of the same order in the determination of the target's $\vec{r}_o$ coordinates, although it will be typically much smaller and decreases rapidly for objects that are further away:  $\Delta \vec{r}_o \approx ({z_0}/{|\vec{r}_0|})\Delta z$.

\paragraph{}As mentioned above, the calculated probability densities $P_i^{\text{ellipse}}$ of each pixel are multiplied to obtain the joint probability density $P(\vec{r}_o)$. However, there is a risk that a given pixel $i$ will lead to an unsignificant fit of the signal and that the values of $\langle t\rangle_i$ and $\sigma_{t_i}$ will be unreliable.  To avoid that these unreliable $P_i^{\text{ellipse}}(\vec{r}_o)$ affect the joint probability density by multiplying relevant densities by zero, we take the probability $P_i(\vec{r}_o)$ associated with the pixel $i$ to be a linear combination of the probability density $P_i^{\text{ellipse}}(\vec{r}_o)$  and a uniform probability density $P^{\text{uniform}}$ that will prevent any point in space to be multiplied by zero,

\begin{align}\label{lin_comb}
P_i(\vec{r}_o) = \alpha_i\ P_i^{\text{ellipse}}(\vec{r}_o) + (1-\alpha_i)P^{\text{uniform}}.
\end{align}

Here, $\alpha_i$ is a coefficient between 0 and 1, related to the reliability of the probability density $P_i(\vec{r}_o)$. Our choice of $\alpha_i$ for each pixel depends on how well the probability density $P_i(\vec{r}_o)$ overlaps with the space in which we are searching the object. More precisely we set $\alpha_i = A_i/A$ where $A_i$ is the area contained in the search space where the probability density $P_i(\vec{r}_o)$ is over a certain threshold (half its maximum) and $A$ is the area of the search space.

\paragraph{}Given a number of pixels $n$, each with associated probability $P_i(\vec{r}_o)$, the joint probability density $P(\vec{r}_o)$ is defined as

\begin{align}\label{stcompress10}
P(\vec{r}_o) = N \prod_{i=1}^n P_i(\vec{r}_o).
\end{align}

where $N$ is a normalisation constant.

\section{Tracking moving objects}

\begin{figure}[h]
\subfigure{\textbf{a}
\includegraphics[width=0.49\textwidth]{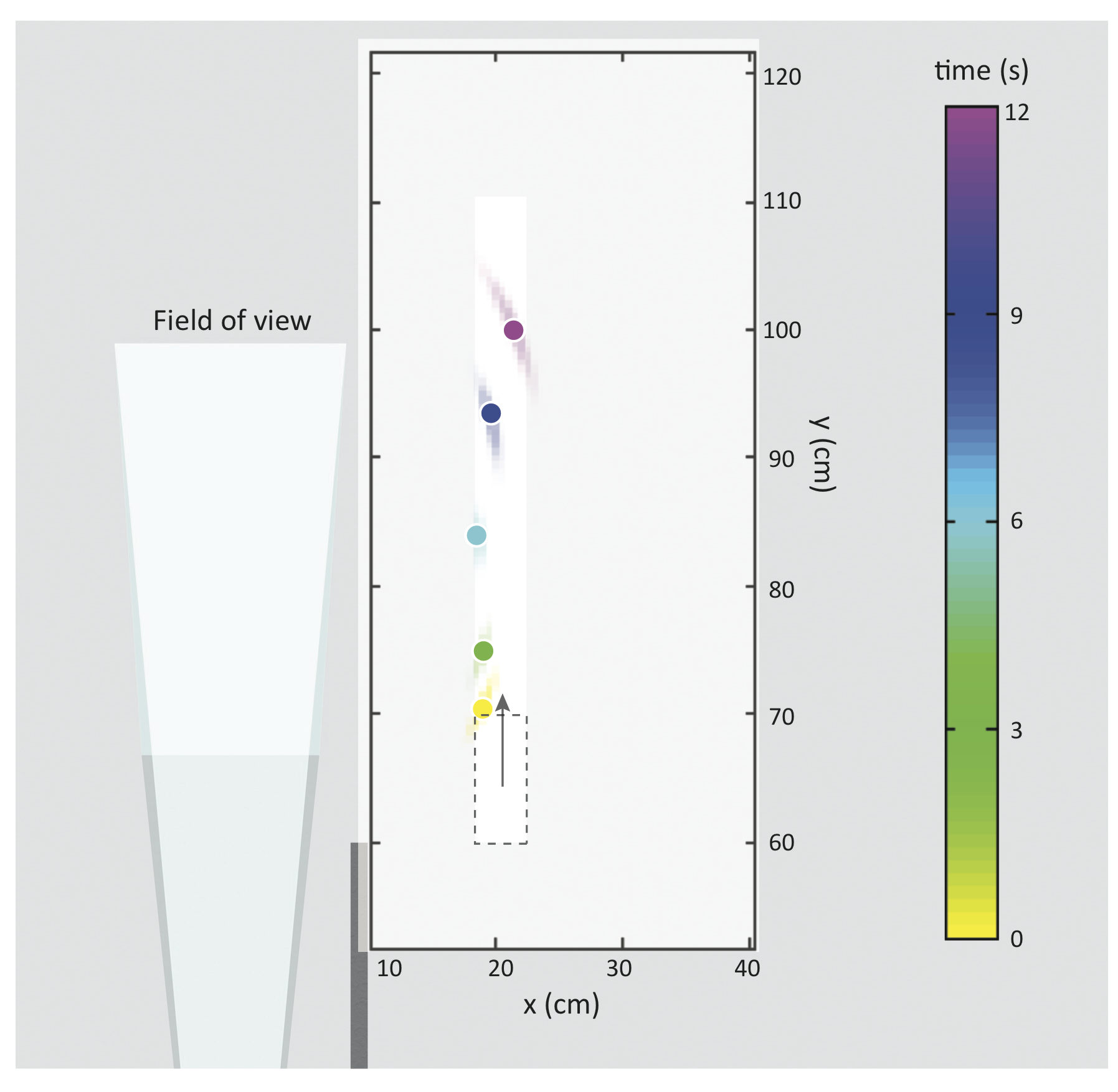}}
\subfigure{\textbf{b}
\includegraphics[width=0.5\textwidth]{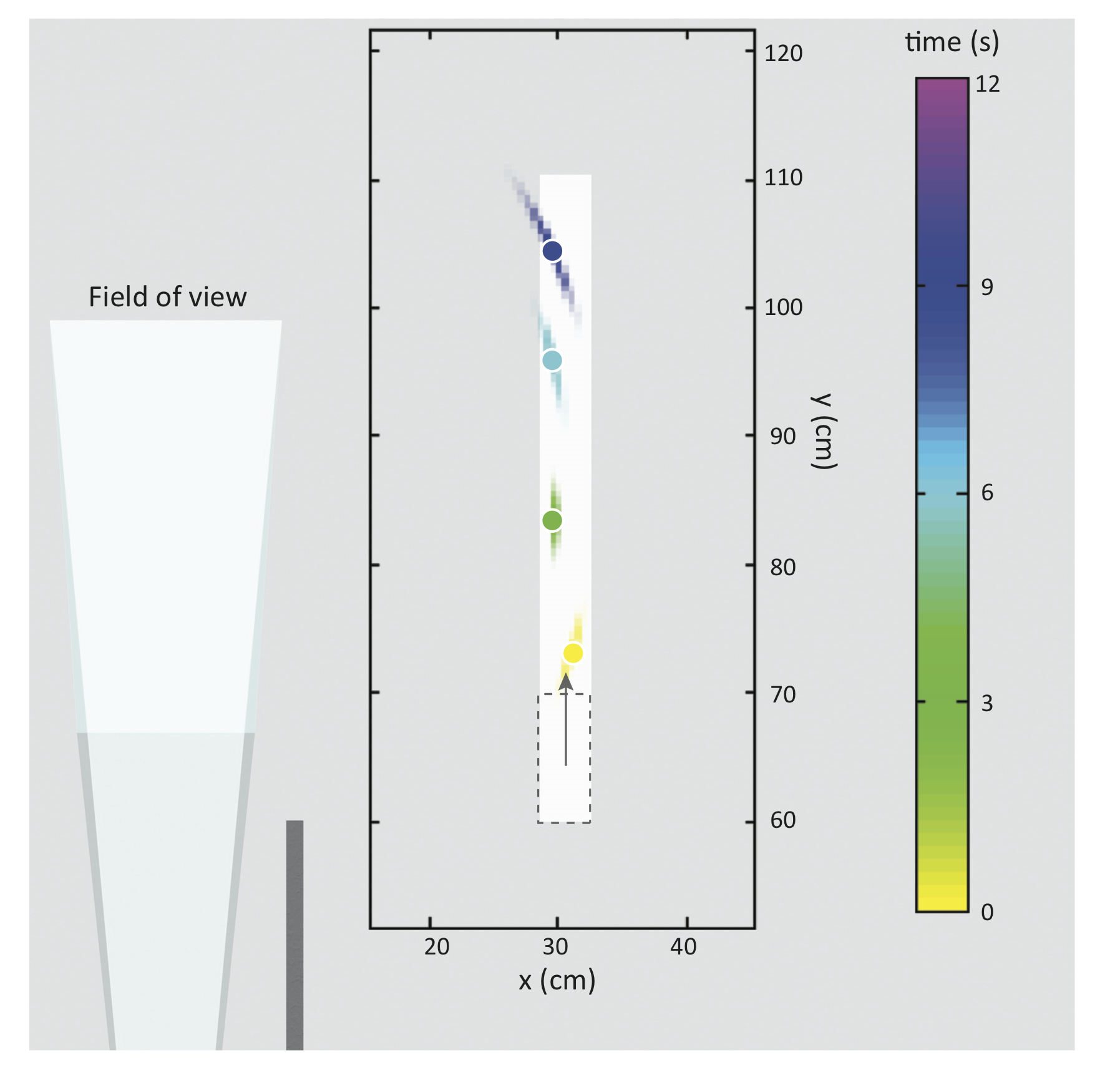}}
\caption{\label{track3}\textbf{Tracking moving object.} The target is moving in a straight line along the $y$-direction, from bottom to top, at a speed of 3.9~cm/s in a) and 5.3~cm/s in b. The retrieved probability density of the target's location are represented in color, where the color indicates the time at which the position is recorded. The acquisition was done in real-time as the object was moving along the tracks.}
\end{figure}

We show in the main text the results for an experiment where the target is moving at 2.8~cm/s on a track positioned at $x=40$ cm. Two other sets of data were acquired, for tracks positioned at $x=20$ cm and $x=30$ cm and the object is moving along the track at 3.9~cm/s and 5.3~cm/s, respectively. We show in Figure \ref{track3} consecutive positions acquired for the target moving along these two other tracks. These results are similar to those shown in Figure 4 of the main text, although they are more precise as the target is closer to the wall. We can also see that the difference in distance covered by the target moving at different speed. 

\end{document}